\documentclass[10pt]{iopart}
\usepackage{iopams}
\expandafter\let\csname equation*\endcsname\relax
\expandafter\let\csname endequation*\endcsname\relax
\usepackage{amsmath}
\usepackage{iopams}
\usepackage{graphicx}
\usepackage{bm,epsfig}
\usepackage[export]{adjustbox}
\usepackage{amssymb}
\usepackage{latexsym}
\usepackage{setspace}
\usepackage{subfigure}

\usepackage[dvips]{pstricks}
\begin{document}

\title{Multi-photon Resonance Phenomena Using Laguerre-Gaussian Beams}

\author{Seyedeh Hamideh Kazemi and Mohammad Mahmoudi}

\address{Department of Physics, University of Zanjan, University Blvd., 45371-38791, Zanjan, Iran}
\ead{mahmoudi@znu.ac.ir}
\vspace{10pt}
\begin{indented}
\item[]{\today}
\end{indented}

\begin{abstract}
We study the influence of laser profile on the linewidth of the optical spectrum of multi-photon resonance phenomena. First, we investigate the dependence of the absorption spectrum on the laser profile in a two-level system. Thanks to the Laguerre-Gaussian field, the linewidth of the one-photon optical pumping and two-photon absorption peaks are explicitly narrower than that obtained with a Gaussian field. In next section, it is shown that, comparing with the the Gaussian fields, the Laguerre-Gaussian ones reduce the linewidth of the optical spectrum in the coherent population trapping. Interestingly, it turns out that its use of a Laguerre-Gaussian beam makes the linewidth of the spectrum narrower as compared with a Gaussian one in a Doppler-broadened electromagnetically induced transparency. Moreover, we study the effect of the laser profile on the Autler-Townes doublet structure in the absorption spectrum for a laser-driven four-level atomic system. We also consider the different values of the Laguerre-Gaussian mode beam waist, and, perhaps more remarkable, we find that for the small values of waist, the Autler-Townes doublet can be removed and a prominent narrow central peak appears in the absorption spectrum. Finally, we investigate the effect of the laser profile on the linewidth of the sub-natural three-photon absorption peak of double dark resonance. The differences in the linewidth are quite large, offering potential applications in metrology and isotope separation methods. Our results can be used for super ultra high resolution laser spectroscopy and to improve the resolution of the technology of isotope/isomer separation and photo-biology even at essential overlap of the spectra of the different particles.
\end{abstract}

%\pacs{42.62.Fi, 42.50.Gy, 32.70.Jz, 32.80.Qk}

%42.62.Fi 	Laser spectroscopy, 32.70.Jz 	Line shapes, widths, and shifts,33.40.+f 	Multiple resonances (including double and higher-order resonance processes, such as double nuclear magnetic resonance, electron double resonance, and microwave optical double resonance),42.50.Gy 	Effects of atomic coherence on propagation, absorption, and amplification of light; electromagnetically induced transparency and absorption, 42.65.An 	Optical susceptibility, hyperpolarizability [see also 33.15.Kr Electric and magnetic moments (and derivatives), polarizability, and magnetic susceptibility], 32.80.Qk 	Coherent control of atomic interactions with photons

% \vspace{2pc}
%\noindent{ Keywords}: Multi-photon resonance, Laguerre-Gaussian beam, Coherent control
%\maketitle
\ioptwocol

\section{Introduction}
Non linear interference effects in the resonant atom-light interaction form the base for a wealth of important phenomena, such as coherent population trapping (CPT) \cite{moi,moi2,moi3,moi4}, lasing without inversion \cite{ya,ya2,ya3,ya4,ya5,ya6,ya7,ya8}, electromagnetically induced transparency (EIT) \cite{harris,harris2} and electromagnetically induced absorption (EIA) \cite{lezama}. Besides the fundamental interest, they have found numerous applications in non linear high-resolution spectroscopy \cite{1,2,3,4,41}, laser cooling \cite{5,6}, atom optics and interferometry \cite{7,8,81}, quantum information processing \cite{9,10,11,12,121} and high-precision metrology (atomic clocks and magnetometers) \cite{21,22,23,24}. Especially one more application should be noted, i.e., isotope/isomer discrimination. This technique is based on this fact that different isotopes/isomers absorb light at different frequencies in order that one isotope may be selectively excited by a laser tuned to a resonant atomic transition, resulting in discriminating from the another isotope. In 1996, Kasapi \cite{kasapi} proposed a laser-based isotope separation method between two isotopes/isomers of the same element using EIT. While one isotope/isomer is made resonantly opaque, another is simultaneously rendered highly transparent to a probe field. This separation method offers broad prospects for various applications because the enhanced signal is about one order of magnitude higher compared with conventional isotope/isomer discrimination methods.

As a rule, an interaction involving a dark state leads to undesirable decoherence, while multiple coherent interacted quantum superposition states that are coupled, i.e., double-dark resonance (DDR) \cite{1feng,7feng}, can be used to reduce the various decoherence effects. In 1999, Lukin \textit {et al} introduced the idea of DDR and showed that in a $\Lambda$- type system, the coherent interaction can lead to a splitting of dark states and the emergence of sharp spectral features \cite{1feng}. Later, it was shown that this novel spectral feature can be characterized as a powerful mechanism for Doppler-free resonance \cite{7feng,6feng}, group velocity controlling \cite{11feng,12feng}, coherently control the adiabatic passage \cite{2feng,3feng}, four-wave mixing \cite{8feng,10feng} and high resolution spectroscopy \cite{garva,chu}. Up to now, different schemes of the DDR have been investigated and also various phenomena induced by the interacting dark resonance have been explored. For example, Goren \textit{et al} studied a sub-Doppler and sub-natural narrowing in a nearly degenerate tripod atomic system \cite{299}. Recently, Verma and Dey proposed a scheme in order to improve the resolution and contrast of diffraction-limited images imprinted onto a probe field, based on the quantum interference effects induced by interacting dark resonances \cite{verma}. It has been also theoretically demonstrated that multi-layered plasmonic meta-molecules exhibit sub-natural spectral response, analogous to conventional atomic four-level configurations \cite{kim}. In these applications, the sub-natural spectral response has a major role in these effects. Thus, it is evident that the narrower peak, the more pronounced  these effect.

On the other hand, Laguerre-Gaussian (LG) light beams \cite{allen,he} having a doughnut-shaped intensity distribution and zero intensity at the beam centre, have given birth to various excellent applications such as creation of a waveguide in an atomic vapour \cite{truscott,kapoor}, rotating trapped microscopic particles \cite{Paterson} and a narrowing of the line shape of the Hanle resonance \cite{hanle}. Recently, an LG beam have been used as a control beam in EIT and EIA, leading to narrow resonances at the line center of an optical transition \cite{sapam}. In another study, Akin \textit{et al} \cite{akin} have compared EIT linewidths with the control beam in an LG mode and that in a Gaussian mode and concluded that using an LG mode, the EIT in an ultra cold gas can result in a narrower resonance feature. However, the influence of the laser profile on the multi-photon resonance phenomena is somewhat less studied.

It is well-known that the tunable laser with narrow width is an essential tool for the laser spectroscopy which has been started using the optical pumping, i.e., one-photon transition. However, this phenomena applies generally a fundamental limit to the resolution due to the natural linewidth and thus can not use to distinguish different isotopes/isomers with essential overlap of the optical spectra. It was shown that a two-photon transition, i.e., CPT makes a possibility of sub-natural laser spectroscopy in atomic and molecular systems \cite{Izmailov2}. It is therefore desirable to  generate the narrower structure in optical spectrum in order to  improve the resolution of the spectroscopy. In this paper, we investigate one-, two- and three-photon resonance line profiles obtained with Gaussian and LG beams and show that the LG beams make the linewidth of these spectra explicitly narrower. The LG-beam-induce narrowing may have several substantial applications such as atomic clocks and magnetometers with higher precision, more efficient isotope separation methods and increased storage times in slow light experiments. In addition, the use of the LG fields may prove advantageous in all applications of laser spectroscopy.

\section{General Formalism }
Laguerre-Gaussian beam ($LG_{p}^{l}$) defines a solution of the paraxial wave function in a cylindrical coordinate where its indices $l$ and $p$ are the number of times the phase completes $ 2 \pi $ on a closed loop around the axis of propagation and the number of radial node for radius $r>0$, respectively \cite{hanle,allen1}.

In this paper we use an $LG_{0}^{1}$ mode, where the electric field associated with the mode is given by
\begin{equation}\label{eq21}
\vec{E}_{LG}=\hat{e} \, E_{0LG}\, \frac{  r \,e^{i \psi} }{w_{LG}} \, \exp(-r^2/w_{LG}^2),
\end{equation}
here $w_{LG}$, $\hat{e}$ and $e^{i \psi}$ denote the beam waist, the unit polarization vector and the reduced mode amplitude of the electric field, respectively. For a Gaussian beam, the electric field is defined as  $\vec{E_{G}}= \hat{e} \, E_{0G}\, \exp( -r^2/ w_{G}^2)$ with $w_{G}$ being beam waist of the field. General expression for a Rabi frequency is defined as $g=( \vec{E} . \vec{\mu})/{\hbar}$ with $\vec{\mu}$ and $\vec{E}$ being the atomic dipole moment of the corresponding transition and the peak amplitude of the field, respectively. For the LG field, the Rabi frequency is $g=g_{0} \, r \,e^{i \psi}  \, \exp( - r^2/ w_{LG}^2) / w_{LG} $ with $g_{0}=(E_{0LG}\,\hat{e}.\vec{\mu})/{\hbar}$ as Rabi frequency constant. The corresponding express for the Gaussian beam is $g=g^{'}_{0}\, \exp( -r^2/w_{G}^2)$ with $g^{'}_{0}=(E_{0G}\,\hat{e}.\vec{\mu})/{\hbar}$.

It is worth to mention that in order to compare different configurations of the LG and Gaussian beams, it is necessary that the fields have the same total laser power, $P= \int  f(r) \, 2 \pi r  \ dr$ with $f(r)$ being the beam radius profile function. So in all figures in which they appear, the field amplitudes are chosen in order that the fields will have the same power. Also, the absorption of each figure has been normalized in order to aid linewidth comparison. For experimental point of view, the LG mode beam waist is smaller than the Gaussian mode beam waist which is necessary to counteract the loss of laser power after converting the Gaussian mode to the LG one. Throughout our discussion of numerical results we assume, unless otherwise stated explicitly, that the field has a  $1/e^2$ radius of $270 \,\mu m$ when in the $LG_{0}^{1}$ mode and  $1.1\, m m$ when in the Gaussian mode \cite{akin}.

In order to study optical properties of each system, we will derive a Hamiltonian of the system and then using the Von Neumann, $ i \hbar \, \frac{d}{dt}\rho$ $= [H,\rho]$, the equations of motion for the considered system can easily be derived.
\section{Two-level System}
\subsection{Optical Pumping}

\begin{figure}
\centering
\includegraphics[width=7 cm]{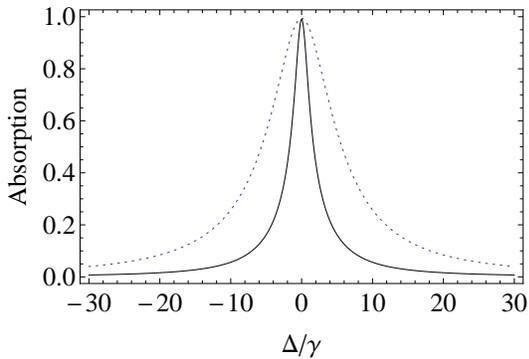}
\caption{ Normalized absorption spectrum versus the detuning $\Delta$ for $T_{1}=T_{2}=1$ and $g_{0}=g^{'}_{0}=5 \gamma$. The solid curve is calculated assuming a laser beam of an $LG^{1}_{0}$ mode, and the dotted one belongs to the field in a Gaussian mode. }
\label{fig1}
\end{figure}

We first consider a closed two-level system in which the transition between upper level $\vert 2 \rangle$ with energy $E_{2}=\hbar \omega_{22}$ and  lower level $\vert 1 \rangle$ with energy $E_{1}=\hbar \omega_{11}$ is driven by a field at frequency $\omega_{l}$ and with Rabi frequency $g$. We represent a Hamiltonian for this system as
\begin{equation}
H= H_{0} + V(t),
\end{equation}
where $H_{0}$ and $V$ denote an atomic and interaction Hamiltonian, respectively. The Hamiltonian $H_{0}$ is represented by a diagonal matrix whose elements are given by $H_{0,nm}=E_{n}\delta_{nm}, (n, m \in \lbrace 1, 2\rbrace)$. We then assume that the interaction energy can be described in the electric dipole approximation so that  the interaction Hamiltonian has the form as
\begin{equation}
V(t)= -\mu E.
\end{equation}
 We assume that the applied field is given by $E(t)= E  \,e^{-i\omega_{l} t } +c.c.$ and that atomic functions have a definite parity and the diagonal matrix elements of $\mu$ vanish and non-vanishing elements of $V$ are given by $V_{21}=V^{*}_{12}=- \mu _{21} E e^{-i\omega_{l} t }$.
Within the rotating-wave approximation, the time evolution of the density matrix is given by \cite{boyd}
\begin{subequations}
\label{eq1}
\begin{eqnarray}
& &\frac{d}{dt}  \rho_{21} =- ( i \bar{\omega}_{21} + \frac{1}{T_{2}} ) \,\rho_{21} \nonumber \\
&-&\frac{i}{\hbar} \mu_{21}\, E \, e^{-i\omega_{l} t }( \rho_{22}- \rho_{11}),  \\
& &\frac{d}{dt}  (\rho_{22}- \rho_{11}) =- \frac{(\rho_{22}-\rho_{11})- (\rho_{22}-\rho_{11})^{(eq)}}{T_{1}} \nonumber \\
&+&  \frac{2 i}{\hbar}( \mu_{21} E \, e^{-i\omega_{l} t }\,\rho_{12}-\mu_{12} E^{*}\, e^{i\omega_{l} t }\rho_{21}),
\end{eqnarray}
\end{subequations}
where we assume that the atomic dipole moment is dephased in the characteristic time $T_{2}$, leading to a transition linewidth of characteristic width $\gamma_{21}=1/T_{2}$. Note that the population inversion, $\rho_{22}- \rho_{11}$, relaxes from its initial value to its equilibrium value $(\rho_{22}-\rho_{11})^{(eq)}$ in a time of the order of $T_{1}=1/ \gamma$ which is called the population relaxation time. Also, the parameter $\bar{\omega}_{21}$ denotes the transition frequency.
We now concentrate on the response of the atomic medium to the field. Linear susceptibility, $\chi=\chi_{1}+i \chi_{2}$ of the field can be written as \cite{scully}
\begin{equation}
\chi=\frac{2N\mu_{12}}{\epsilon_{0}\, E}  \rho_{21},
\end{equation}
where $N$ is the atom number density in the medium.
To investigate the absorption spectrum, we proceed by finding the steady-state coherence $\rho_{21}$ from equations (4) for spatially-dependent Rabi frequency, either $LG^{1}_{0}$ or Gaussian, leading to a radial dependence of the absorption. Here, we define the quantity $g= \mu_{21} E/ \hbar $ which is known as Rabi frequency. Normalized absorption spectrum as a function of the detuning $\Delta=\omega_{l}-\bar{\omega}_{21}$ for $T_{1}=T_{2}=1$ and $g_{0}=g^{'}_{0}=5 \gamma$ is shown in figure \ref{fig1}. Solid line (dotted line) shows the spectrum when the field is in an $LG^{1}_{0}$ (a Gaussian) mode. It is seen from the figure that the $LG^{1}_{0}$ beam causes a narrowing of the spectrum compared to the usual Gaussian beam.

\begin{figure}
\centering
\includegraphics[width=7 cm]{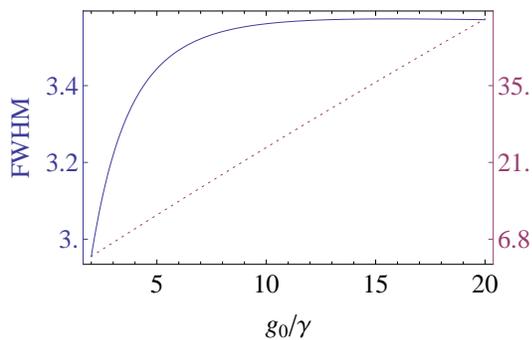}
\caption{The linewidths of the profile of a $LG^{1}_{0}$ beam (solid line) and that of a Gaussian beam (dotted line) versus Rabi frequency constants.}
\label{fig1.1}
\end{figure}

In the following, we are going to compare the linewidths of the above structures, which are determined by the full width at half-maximum (FWHM). Figure \ref{fig1.1} shows the calculated linewidths of the profile as a function of Rabi frequency constant. The figure shows that for all values of the Rabi frequency constant, the absorption linewidth for $LG^{1}_{0}$ mode is smaller than that with the Gaussian field.

\subsection{Two-level Pumped-probe System}
\begin{figure}
\centering
\includegraphics[width=5 cm]{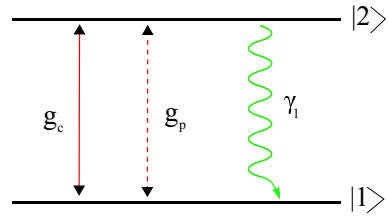}
\caption{ Schematic energy diagram of a two-level atom in the simultaneous presence of the pump field (solid) and probe one (dashed). Wavy lines show the decay rate from the upper level to the lower one.}
\label{figb}
\end{figure}

We then consider another common type two-level system, i.e., a pumped-probe system and investigate the response of the two-level atom to the simultaneous presence of a strong optical field and a weak optical field. A schematic containing the essential features of our pump-probe system is shown in figure \ref{figb}, where the atom with a ground state $\vert 1 \rangle$ and an excited state $\vert 2 \rangle$ is driven by a pump field at frequency $\omega_{c}$ and weak probe field at frequency $\omega_{p}$.
In the simultaneous presence of the pump and the probe field, matrix element of the interaction Hamiltonian is given by
\begin{equation}
 V_{21}(t)=-\mu_{21} (E_{0}+E_{p} e^{-i \delta t}) \, e^{-i \omega_{c} t} +h.c.
\end{equation}\label{eq3}
We represent the amplitude of the field $E=E_{0}+E_{p} e^{-i \delta t}$ with $ E_{0}$ and $E_{p}$ being the amplitudes of the pump and probe fields, respectively. Parameter $\mu_{21}$ is the dipole moment and $\delta= \omega_{p}- \omega_{c}$ denotes the pump field detuning with respect to the probe field.
The density matrix equations of motion are given explicitly by \cite{scully}
\begin{subequations}
\begin{eqnarray}
\frac{d}{dt}  \rho_{22} &=&-\gamma_{1} \rho_{22}+\frac{i}{\hbar} ( V_{21} \rho_{12}- \rho_{21} V_{12}),  \\
\frac{d}{dt}  \rho_{11} &=& \gamma_{1} \rho_{11}-\frac{i}{\hbar}  ( V_{21} \rho_{12}-  \rho_{21} V_{12}), \\
\frac{d}{dt}  \rho_{21} &=&- (i \bar{\omega}_{21}+ \gamma_{2}) \rho_{21}+\frac{i}{\hbar}  V_{21}\, (\rho_{22}- \rho_{11}),
\end{eqnarray}\label{eq4}
\end{subequations}
where we have introduced the transition frequency $\bar{\omega}_{21}=(E_{2}-E_{1})/\hbar$ with $E_{i}$ being the energy of non-perturbed state of the system. Similarly, the decay rate from the upper level to the lower level is given by $\gamma_{1}$ and therefore the lifetime of the upper level is $T_{1} = 1/	\gamma_{1}$. We also assume that the atomic dipole moment is dephased in the characteristic time $T_2$, leading to a transition linewidth of characteristic width $\gamma _{2}=1/T_{2}$.

In order to find the steady-state solutions to equations (\ref{eq4}), we proceed to use rotating frames as $\rho_{21}=\tilde\rho_{21} \, e^{-i \omega_{c} t}$, $\rho_{22}=\tilde\rho_{22}$ and $\rho_{11}=\tilde\rho_{11}$, and present the density matrix equations as
\begin{subequations}
\begin{eqnarray}
\frac{d}{dt}  \tilde\rho_{22} &=&-  \gamma_{1}\, \tilde\rho_{22}+i (g_{c}+g_{p} e^{-i \delta t} )\, \tilde\rho_{12} \nonumber \\
&-&i (g_{c}+g_{p} e^{i \delta t})\, \tilde\rho_{21} ,  \\
\frac{d}{dt}  \tilde\rho_{11} &=&  \, \gamma_{1} \tilde\rho_{11}-i (g_{c}+g_{p} e^{-i \delta t} )\, \tilde\rho_{12}\nonumber \\
&+&i (g_{c}+g_{p} e^{i \delta t} )\, \tilde\rho_{21} , \\
\frac{d}{dt}  \tilde\rho_{21} &=& (i \Delta- \gamma_{2}) \tilde\rho_{21}\nonumber \\
&-&i (g_{c}+g_{p} e^{-i \delta t}) (\tilde\rho_{22}- \tilde\rho_{11}).
\end{eqnarray}\label{eq5}
\end{subequations}
We define the Rabi frequencies for the probe and pump fields by $g_{p}=\mu _{21} E_{p}/\hbar$ and $g_{c}=\mu _{21} E_{0}/\hbar$, respectively. The parameter $\Delta=\omega_{c} - \bar{\omega}_{21}$ denotes the pump field detuning from the atomic resonance transition.

\begin{figure}
\centering
\includegraphics[width=7 cm]{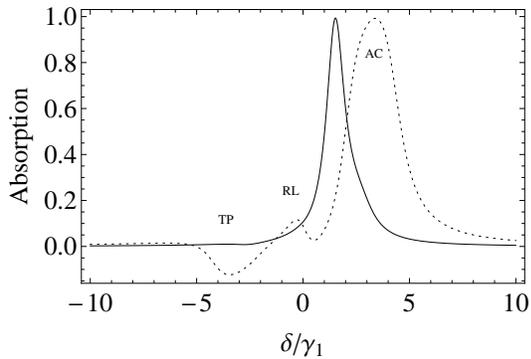}
\caption{ Normalized absorption spectrum of a probe field in the presence of a strong pump field versus the detuning $\delta$ for $\Delta T_{2}=3$, $T_{2}/T_{1}=2$ and $ g_{0} = g_{0}^{'} =4/ T_{2}$. Solid curve is calculated assuming the laser beams of an $LG^{1}_{0}$ mode, and the dotted one belongs to the fields in the Gaussian modes. Also, TP, RL and AC denote the three-photon resonance, the Rayleigh resonance and the Ac-Stark-shifted atomic resonance \cite{boyd}.}
\label{fig2}
\end{figure}
Due to the explicit time dependence of equations (\ref{eq5}), we cannot readily solve exactly for the field, i.e., $E=E_{0}+E_{p} e^{-i \delta t}$. So a strategy will be to find a solution that is exact in the amplitude of the pump field and is correct to lowest order in the amplitude of the probe field, leading to the following solution for an effective linear susceptibility $\chi_{eff}$  \cite{boyd}
\begin{eqnarray}
&&\chi_{eff}= \frac{N \, \vert \mu_{21} \vert ^2\, \hbar\, w_{0}}{\epsilon_{0} \, \hslash \, D(\delta)} \nonumber \\
 &&\times [ ( \delta + \frac{i}{T_1})(\delta + \Delta + \frac{i}{T_2})  +2 g^{2} \frac{\delta}{\Delta + iT_{2}}].
\end{eqnarray}
Here, we have introduced the Rabi frequency $g= \vert  \mu_{21} E\vert / \hbar$ and $w_{0}= (1+ \Delta^{2}\, T^{2}_{2}) /(1+ \Delta^{2}\, T^{2}_{2} + g^{2} \,T_{2} \,T_{1} )$. The denominator of this expression is given by $D(\delta) =(\delta + i/T_1)(\delta + \Delta + i/T_{2})(\delta - \Delta+i/T_{2})- g^{2}\,(\delta+ i/T_{2})$.

Now, we investigate the absorption of the probe field. In figure \ref{fig2}, we display the absorptive response for the case that pump field is detuned from the atomic transition by three linewidths, $\Delta \, T_{2}=3$, and $T_{2}/T_{1}=2$. Solid line (dotted line) in figure \ref{fig2} shows the absorption of the probe field when the fields are in the $LG^{1}_{0}$ (Gaussian) modes. Similarly to the case of the optical pumping, we can achieve a narrower peak using the $LG^{1}_{0}$ fields.

\section{Three-level System}
\subsection{Coherent Population Trapping}
In the present section, we investigate the effect of the laser profile on the CPT. At this phenomenon, the whole population is accumulated to a certain quantum superposition state of a multi-level quantum system subject to decay processes, which is decoupled from the multi-component laser radiation. Such trapping of population is by now a well-known concept in quantum optics and laser spectroscopy, forming the basis of a number of prominent applications such as ultra high resolution spectroscopy, atomic clocks, magnetometry and coherent population transfer among quantum states of atoms/molecules, the review articles by Arimondo \cite{Arimondo2} and Aga\`{p}ev \textit{et al} \cite{Agapev} may be referred. Thus, the narrower peaks in the populations could prove advantageous in the all application of the CPT.
\begin{figure}
\centering
\includegraphics[width=5 cm]{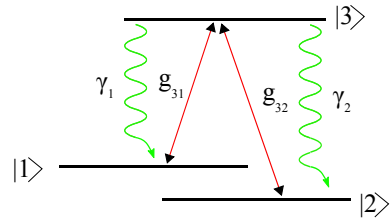}
\caption{ Schematic diagram of a three-level $\Lambda$-type system of optical transitions $\vert 1\rangle - \vert 3\rangle$ and $\vert 2\rangle - \vert 3\rangle$ between the excited level $\vert 3\rangle$ and the states  $\vert 1\rangle$  and $\vert 2\rangle$.}
\label{figc}
\end{figure}
In spite of the large activities in the field of the CPT in closed atomic systems, this phenomenon in molecular multi-level systems, in which the levels involved in the CPT do not form a closed system, has not yet been studied exhaustively. In fact, because of the richness of molecular excitation pathways and variety of molecules, it is becoming increasingly more possible to develop novel high resolution spectroscopic applications of various quantum optics tools. So, we consider an interaction of molecules of a gas on level of an open $\Lambda$-system with two monochromatic light waves: a basic model for a wide array of excitation processes in molecular and solid-state systems (figure \ref{figc}).
Electric component of the field is given by
\begin{equation}
\vec{E}= \sum_{i=1}^{2} \hat{e}_{i}\, E_{0}\,  \exp [i(\omega_{i} t - \vec{K}_{i}. \vec{r})]+ c.c.,
\end{equation}
 where $\hat{e}_{i}$, $E_{0} (r)$, $\omega_{i}$ and $\vec{K}_{i}$ are the unit polarization vector, the amplitude of the field, the frequency and the wave vector, respectively. Noting that central frequency is defined as $\bar{\omega}_{3i}=(E_{3}-E_{i})/\hbar$ with $E_{i}$ being the energy of non-perturbed state of the $\Lambda$-molecule.

\begin{figure}[h]
\centering
\includegraphics[width=7 cm]{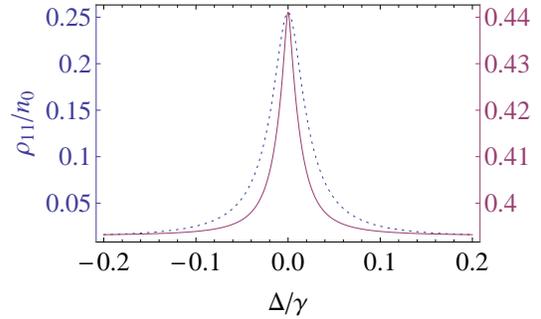}
\caption{Stationary population of level $\vert 1 \rangle$ for the fields in the $LG^{1}_{0}$ modes (solid) and that for Gaussian mode fields (dotted) versus the detuning $\Delta$, when $\gamma_3=\gamma$,  $\Gamma_{12}=\Gamma = 0.002 \, \gamma$, $\rho^{(0)}_{11}=\rho^{(0)}_{22}=0.5 \, n_{0}$, $\Delta_{1}=0$, $\phi_{1}=\phi_{2}=0$, $g_{01}=g^{'}_{01}=g_{02}=g^{'}_{02}=0.4$ and $(\gamma_1+\gamma_2) \ll \gamma$.}
\label{fig3}
\end{figure}

Using the Von Neumann equation for the density matrix, $ i \hbar \, \frac{d}{dt}\rho$ $= [H,\rho]$, and taking into account characteristic relaxation processes in the gas medium, the equations of motion for the considered system can easily be derived to give \cite{Rautian}
\allowdisplaybreaks
\begin{subequations}
\begin{eqnarray}
\frac{d}{dt}  \rho_{11} &=& i \,g_{1}\, \rho_{31} \exp{(i f_1)}+c.c.+2\, \gamma_{1}\, \rho_{33}\nonumber \\
&-& (\rho_{11} -\rho^{(0)}_{11} )\, \Gamma_{1}, \\
\frac{d}{dt}  \rho_{22} &=& i \,g_{2} \, \rho_{32} \exp{(i f_2)}+c.c.+2 \,\gamma_{2} \, \rho_{33}\nonumber \\
&-& (\rho_{22} -\rho^{(0)}_{22} )\, \Gamma_{2}, \\
\frac{d}{dt}  \rho_{33} &=&  - i\, g_{1}\, \rho_{31} \exp{(i f_1)} - i \,g_{2}\, \rho_{32} \exp{(i f_2)} \nonumber \\
&+&c.c\, -2\,  \gamma_{3}\, \rho_{33}\\
\frac{d}{dt}  \rho_{13} &=&   i\, g_{1} (\rho_{33}-\rho_{11}) \exp{(i f_1)} - i \,g_{2} \,\rho_{12} \exp{(i f_2)}\nonumber \\
&-&  \gamma_{13} \, \rho_{13} , \\
\frac{d}{dt}  \rho_{23} &=&   i \, g_{2} (\rho_{33}-\rho_{22}) \exp{(i f_2)} - i \, g_{1} \rho_{21} \, \exp{(i f_1)}\nonumber \\
&-& \, \gamma_{23} \rho_{23} , \\
\frac{d}{dt}  \rho_{12} &=&   i\, g_{1} \rho_{32}\exp{(i f_1)} - i\, g_{2} \rho_{13} \exp{(-i f_2)}\nonumber \\
&-&  \Gamma_{12}\, \rho_{12}.
\end{eqnarray}\label{eq6}
\end{subequations}
Here, $f_{i}= \Delta_{i} t - \vec{K}_{i}. \vec{r} + \phi_{i}$ with $\phi_{i}$ being the argument of the complex Rabi frequency. Parameter $ \Delta_{i}= (\omega_{i}- \bar{\omega}_{3i})$ is the frequency detuning from the resonance for the field $i$. Relaxation rate of the excited state, partial rate of the radiative decay from the excited level to the state $\vert 1\rangle$ and that of to the state $\vert 2 \rangle$ are denoted by  $2 \gamma_{3}$, $2 \gamma_{1}$ and  $2 \gamma_{2}$, respectively. The value $\Gamma_{i}$ describes the relaxation of the population of the long-lived $\vert i\rangle$ to the equilibrium $\rho^{(0)}_{ii}$ because of molecular collisions. We also define the relaxation rate of the light-induced coherence between states $\vert 1\rangle$ and $\vert 2\rangle$ as $\Gamma_{12}$. Moreover, $\gamma_{13}$ and $\gamma_{23}$ are half-width of spectral for transition $\vert 1\rangle - \vert 3\rangle$ and $\vert 2\rangle - \vert 3\rangle$, respectively. We further assumed the constraint, $\rho_{ij}=\rho^{*}_{ji}$.
In the following, we use replacements for non-diagonal elements of the density matrix in equations (\ref{eq6})
\begin{subequations}
\begin{eqnarray}
 \rho_{i3}(t) &=&   \tilde \rho_{i3}(t) \exp{(i[\Delta_{i} \, t -\vec{K}_{i} . \vec{r}])} \,\,\,\,\,\,\ (i=1,2),\\
\rho_{12}(t) &=&   \tilde \rho_{12}(t) \exp{(i[(\Delta_ 1-\Delta_ 2) t -(\vec{K}_{1}- \vec{K}_{2}). \vec{r}])}.
\end{eqnarray}
\end{subequations}
Then the linear equations follow as \cite{Izmailov2}
\begin{subequations}
\begin{eqnarray}
\frac{d}{dt}  \rho_{11} &=& i G_{1} \, \tilde{\rho}_{31} -i G^{*}_{1} \, \tilde{\rho}_{13} +2 \gamma_{1}\, \rho_{33}\nonumber \\
&-& (\rho_{11} -\rho^{(0)}_{11} )\, \Gamma_{1}, \\
\frac{d}{dt}  \rho_{22} &=& i G_{2} \, \tilde{\rho}_{32} -i G^{*}_{2}\, \tilde{\rho}_{23}+2 \gamma_{2} \, \rho_{33}\nonumber \\
&-& (\rho_{22} -\rho^{(0)}_{22} ) \, \Gamma_{2}, \\
\frac{d}{dt}  \rho_{33} &=&  - i G_{1}\, \tilde{\rho}_{31}  - i G_{2}\, \tilde{\rho}_{32} \nonumber \\
&+& i G^{*}_{1}\, \tilde{\rho}_{13} +  i G^{*}_{2}\, \tilde{\rho}_{23}  -2  \gamma_{3}\, \rho_{33}, \\
\frac{d}{dt}  \tilde{\rho}_{13} &=&  i G_{1} \,(\rho_{33}-\rho_{11})  - i G_{2} \tilde{\rho}_{12} \nonumber \\
&-&  (\gamma_{13}+i \Delta_{1}) \, \tilde{\rho}_{13} , \\
\frac{d}{dt}  \tilde{\rho}_{23} &=&   i G_{2}\, (\rho_{33}-\rho_{22}) - i G_{1} \tilde{\rho}_{21} \nonumber \\
&-&  (\gamma_{23} + i \Delta_{2}) \tilde{\rho}_{23} , \\
\frac{d}{dt}  \rho_{12} &=&   i G_{1} \, \tilde{\rho}_{32} - i G^{*}_{2} \, \tilde{\rho}_{13} \nonumber \\
&-&  (\Gamma_{12} + i (\Delta_{1} - \Delta_ {2}))\, \tilde{\rho}_{12}.
\end{eqnarray}\label{eq7}
\end{subequations}
Here, we define the Rabi frequency $G_{i}= g_{i} \exp{(i \phi_{i})}$.
Considering the restrictions and assuming $\gamma_{13}=\gamma_{23}=\gamma$, $\Gamma_1=\Gamma_2=\Gamma$ and $ \vert g_{p1}\vert=\vert g_{p2}\vert=\vert g_{p}\vert $, the stationary populations  $\rho_{11}$, $\rho_{22}$ and $\rho_{33}$ are given by

\begin{subequations}
\begin{eqnarray}
\rho_{11}&=&\frac{0.5 \, \Gamma (1+g^{2}_{2} \, g^{-2}_{1}) \,(\rho^{(0)}_{11} \, g_{2}^{2} +\rho^{(0)}_{22} \, g_{1}^{2})}{ \gamma (0.25  \, W^{2} + \Delta ^2) },\\
\rho_{22}&=& \frac{g^{2}_{1}}{g^{2}_{2}}\, \rho_{11},
\end{eqnarray}\label{eq8}
\end{subequations}
where $\Delta =\Delta_{2}-\Delta_{1}$ and the characteristic width $W_{0}$ is given by
\begin{equation}
W_{0}= 2 \sqrt{(g^{2}_{1} +  g^{2}_{2}) [\frac{\Gamma_{12}}{\gamma} + 0.5 \,   (\frac{g^{2}_{1}}{g^{2}_{2}} + \frac{g^{2}_{2}}{g^{2}_{1}}) \, \frac{\Gamma}{\gamma}]},
\end{equation}
for the case of the fields in the Gaussian mode, we find a very good agreement of our numerical calculated FWHM and the results obtained by integrating $(\int  W_{0} \, \exp(-r^2/w^{2}_{G}) dr) /w_{G} $.
 Let us analyse the stationary population of the considered system by presenting the results of equations (\ref{eq8}). Figure \ref{fig3} shows the dependency of the population on the frequency detuning of the fields for different profiles. Parameters used are $\gamma_3=\gamma$, $\Gamma_{12}=\Gamma = 0.002 \, \gamma$, $\rho^{(0)}_{11}=\rho^{(0)}_{22}=0.5 \, n_{0}$, $\Delta_{1}=0$,  $\phi_{1}=\phi_{2}=0$ and $g_{01}=g^{'}_{01}=g_{02}=g^{'}_{02}=0.4$. Solid line (dotted line) shows the population when the fields in the $LG^{1}_{0}$ (Gaussian) modes. Thanks to the CPT, narrow peaks appear in the population of the lower state with the center at the point $\Delta=0$ for both profiles, yet the $LG^{1}_{0}$ fields reduce the linewidth of the corresponding spectrum. This LG beam-induce narrowing should be useful, not only for ultra high resolution spectroscopy, but also for technology of the isotope/isomer separation.
\subsection{Doppler broadened Electromagnetically Induced Transparency}
\begin{figure}
\centering
\includegraphics[width=7 cm]{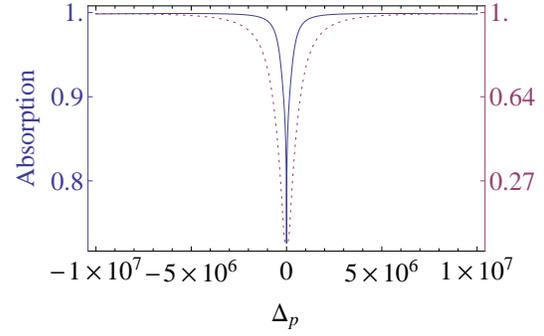}
\caption{Normalized absorption as a function of the probe detuning for the coupling field in the $LG^{1}_{0}$ mode (solid line) and that in the Gaussian mode (dotted line) with integrating the Doppler velocity group for CTP configuration with $\gamma_{1}=\gamma_{2}=\gamma=0.5 MHz$, $g_{p}=0.01 \gamma$, $\Delta_{c}=0$ and $g_{0c}=g^{'}_{0c}= \gamma$. }
\label{fig7}
\end{figure}

Next, we consider a three-level $\Lambda$-type system driven by the strong pump and weak probe fields (figure \ref{figc}). The probe laser is scanned across the $\vert 3\rangle-\vert 2\rangle$ transition with the detuning $\Delta_{p}=\omega_{p}-\bar{\omega}_{32}$ and Rabi frequency $g_{p}=g_{32}$, while the coupling laser drives the $\vert 3\rangle -\vert 1\rangle$ transition with Rabi frequency $g_{c}=g_{31}$ and detuning $\Delta_{2}=\omega_{c}-\bar{\omega}_{31}$. The decay rates from level $\vert 3\rangle$ to the ground levels $\vert 1\rangle$ and $\vert 2\rangle$ are given by $\gamma_1$ and $\gamma_2$, respectively.
The density matrix equations of motion in the rotating wave approximation and in the rotating frame are \cite{mahmoudi8}
\begin{subequations}
\begin{eqnarray}
\frac{d}{dt}  \rho_{11} &=& i g_{c} \, \rho_{31} -i g^{*}_{c} \, \rho_{13} +2 \gamma_{1}\, \rho_{33}, \\
\frac{d}{dt}  \rho_{22} &=& i g_{p} \, \rho_{32} -i g^{*}_{p} \, \rho_{23} +2 \gamma_{2}\, \rho_{33}, \\
\frac{d}{dt} \rho_{12} &=& - i (\Delta_{c}-\Delta_{p}) \rho_{12}+ i g_{c} \, \rho_{32}-i g^{*}_{p} \, \rho_{13} , \\
\frac{d}{dt} \rho_{31} &=&  i (\Delta_{c}-(\gamma_{1}+\gamma_{2})) \rho_{31} \nonumber \\
&+& i g^{*}_{p} \, \rho_{21}- i g^{*}_{c} (\rho_{33}-\rho_{11}) , \\
\frac{d}{dt} \rho_{32} &=&  i (\Delta_{p}-(\gamma_{1}+\gamma_{2})) \rho_{32} \nonumber \\
&+& i g^{*}_{c} \, \rho_{12}- i g^{*}_{p} (\rho_{33}-\rho_{22}).
\end{eqnarray}\label{eq7}
\end{subequations}
More specifically, we are interested in studying the response of the atomic system to the probe field using the susceptibility. The above analysis is correct only for a stationary atom, while for an atom moving along the direction of the beam with velocity $v$, detuning for both probe and coupling lasers will change. For a counter-propagation (CTP) configuration, the following transformations are performed: $\Delta_{p}\rightarrow \Delta_{p}+\omega_{p}v/c$ and $\Delta_{c} \rightarrow \Delta_{c}-\omega_{c}v/c$. Thus for obtaining the probe response for moving atoms, we shall consider the corrected detuning and average the result for the corrected susceptibility over the one-dimensional Maxwell-Boltzmann velocity distribution
\begin{equation}
f(v)= \frac{N}{u \sqrt{\pi}} e^{-v^{2}/u^{2}}.
\end{equation}
Here, the distribution is characterized by the most probable velocity of the atoms $u=\sqrt{2\, K_{B}\,T/ m}$, where $k_{B}$, $T$ and $m$ denote the Boltzmann constant and temperature and mass of the atoms, respectively.
In the following, we discuss the influence of the laser profile on the EIT spectrum, using the $LG^{1}_{0}$ and Gaussian fields rather than the plane-wave ones. The results of such a calculation for room-temperature $Rb$ atoms \cite{iftiquar} with $\gamma_{1}=\gamma_{2}=\gamma=0.5 MHz$, $g_{p}=0.01 \gamma$ and $\Delta_{c}=0$ are shown in figure \ref{fig7}. The solid line (dotted line) presents the result when the coupling field is in an $LG^{1}_{0}$ (a Gaussian) mode. As we expect, the linewidth of the EIT dip becomes narrower when the coupling field is in the $LG^{1}_{0}$ mode compared with the Gaussian field. So, one might expect that its use of the $LG^{1}_{0}$ field will be preferable in laser spectroscopy.
\section{Autler-Townes doublet }
\begin{figure}[h]
\centering
\includegraphics[width=5 cm]{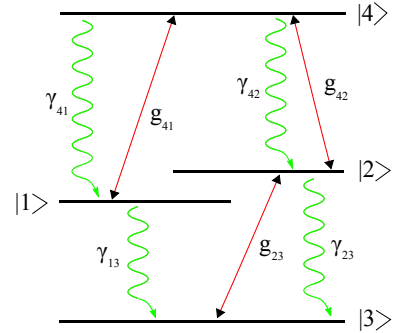}
\caption{Considered energy scheme of a four-level atomic system. Wavy lines show the spontaneous decays from the excited states.}
\label{figd}
\end{figure}

In the following we consider a four-level atomic system, as shown in figure \ref{figd}. A strong coherent field with frequency $\omega_{42}$ is applied to transition $ \vert 2\rangle - \vert 4\rangle$ with Rabi frequency $g_{42}$. A weak coupling field with Rabi frequency $g_{41}$ and frequency $\omega_{41}$ couples to the transition $ \vert 1\rangle - \vert 4\rangle$. Moreover, $ \vert 2\rangle - \vert 3\rangle$ transition is driven by a weak probe field with frequency $\omega_{23}$ and Rabi frequency $g_{23}=g_p$. The spontaneous decay rates are denoted by $ \gamma_{41}$, $ \gamma_{42}$, $ \gamma_{23}$ and $ \gamma_{13}$. The energies of the involved states and the transition frequencies are denoted by $ E_k\, (k \in \lbrace1, ..., 4\rbrace)$ and  $\bar{\omega}_{ij}= (E_{i}-E_{j})/ \hbar \,(i,j \in \lbrace1, ..., 4\rbrace) $, respectively. Moreover, we define detuning of the laser fields as $\Delta_{ij}= \omega_{ij}-\bar{\omega}_{ij}$. Noting that this model has been proposed previously in a different context for group velocity control \cite{12feng}.

The density matrix equations of motion for the considered four-level system are given by \cite{12feng}
\begin{subequations}
\begin{eqnarray}
\frac{d}{dt}  \rho_{11} &=&- i g_{41}^{*} \rho_{14}+ i g_{41} \rho_{42}+2 \gamma_{41} \rho_{44}- 2 \gamma_{13}\rho_{11},  \\
\frac{d}{dt}  \rho_{22} &=& i g_{p}^{*} \rho_{32}-i g_{p} \rho_{23}-i g_{42}^{*} \rho_{24}+ i g_{42} \rho_{42} \nonumber \\
&-& 2 \gamma_{23} \rho_{22}+ 2 \gamma_{42}\rho_{44},\\
\frac{d}{dt}  \rho_{33} &=&- i g_{p}^{*} \rho_{32}+ i g_{p}\rho_{23}+2 \gamma_{13} \rho_{11} +2 \gamma_{23} \rho_{22}\\
\frac{d}{dt}  \rho_{12}&=&-( i\Delta_{41}-i\Delta_{42} + \Gamma_{12})\rho_{12} - i g_{42}^{*} \rho_{14}  \nonumber \\
&+& i g_{41} \rho_{42}- i g_{p} \rho_{13},\\
\frac{d}{dt}  \rho_{13}&=&-( i\Delta_{41}-i\Delta_{42} -i \Delta_{p}+ \Gamma_{13})\rho_{13} - i g_{p}^{*} \rho_{12}\nonumber \\
&+& i g_{41} \rho_{43},\\
\frac{d}{dt}  \rho_{14}&=&-( i\Delta_{41}+ \Gamma_{14})\rho_{14} - i g_{41} \rho_{11}+ i g_{41} \rho_{44}\nonumber \\
&-& i g_{42} \rho_{12},\\
\frac{d}{dt}  \rho_{23}&=&-(- i\Delta_{p}+ \Gamma_{23})\rho_{23} - i g_{p}^{*} \rho_{22}+ i g_{42} \rho_{43}\nonumber \\
&+& i g_{p}^{*} \rho_{33},\\
\frac{d}{dt}  \rho_{24}&=&-(i\Delta_{p}+\Gamma_{24})\rho_{24}+ i g_{p}^{*} \rho_{34}+ i g_{42} \rho_{44}\nonumber \\
&-&i g_{41} \rho_{21}- i g_{42} \rho_{22},\\
\frac{d}{dt}  \rho_{34}&=&-(i\Delta_{p}+i\Delta_{42}+\Gamma_{34})\rho_{34}+ i g_{p} \rho_{24}\nonumber \\
 &-&i g_{42} \rho_{32}- i g_{41} \rho_{31}.
\end{eqnarray}\label{eq91}
\end{subequations}
It ought to be mentioned that the remaining equations follow from the constraints $\rho_{ij}=\rho^{*}_{ji}$ and $\sum _{i} \rho_{ii}=1 $. Moreover, we have defined $\Gamma_{ij}= (2 \gamma _i+ 2 \gamma_ j)/ 2 $ as the damping rate of the coherence with $\gamma_ i$ being the total decay rate out of state $\vert i\rangle$, $\Delta_{p}=\Delta_{23}$ as the probe field detuning and $ \gamma _{j}= \gamma_{1j}+\gamma_{2j}$.
\begin{figure}[h]
\includegraphics[width=7 cm]{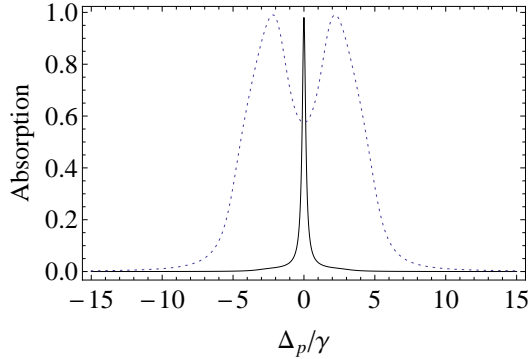}
\caption{ Normalized absorption as a function of the probe detuning $\Delta_{p}$ for $\gamma_{41}=\gamma$,  $\gamma_{13}=0.01 \, \gamma$, $\gamma_{23}=0.14 \gamma$, $\gamma_{42}=0.79 \gamma$, $g_{p}=10^{-4} \gamma$, $\Delta_{42}=\Delta_{41}=0$, $g_{041}=0$ and $g_{042}=g^{'}_{042}=5 \, \gamma$. The solid curve is obtained for the strong coupling field in an $LG^{1}_{0}$ mode, and the dotted curve represents the corresponding result for the field of the Gaussian mode.}
\label{fig4.1}
\end{figure}

We then derive expressions for the linear susceptibility of the weak probe field as our main observable. The normalized absorption as a function of the probe detuning $\Delta_{p}$ for the $LG^{1}_{0}$ and Gaussian beams is shown in figure \ref{fig4.1}. The solid line (dotted line) is for the case of the strong coupling field in an $LG^{1}_{0}$ (a Gaussian) mode. Used parameters are $\gamma_{41}=\gamma$,  $\gamma_{13}=0.01 \, \gamma$, $\gamma_{23}=0.14 \gamma$, $\gamma_{42}=0.79 \gamma$, $g_{p}=10^{-4} \gamma$, $\Delta_{42}=\Delta_{41}=0$ and $g_{042}=g^{'}_{042}=5 \, \gamma$ and $g_{041}=0$. For the case of the strong coupling field in a Gaussian mode, the absorption spectrum of the weak probe laser shows the familiar Autler-Townes doublet structure. More remarkably, the Autler-Townes doublet structure switches to a significantly narrowed peak in the absorption spectrum, when the strong coupling field is in an $LG^{1}_{0}$ mode.

\begin{figure}[h]
\centering
\includegraphics[width=7 cm]{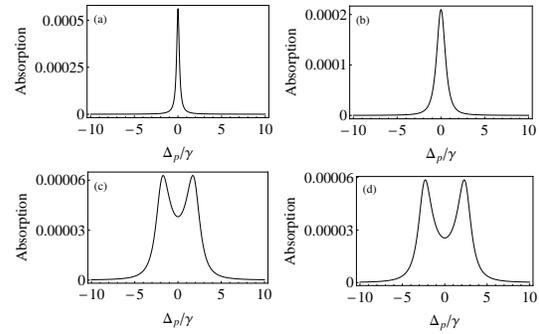}
\caption{Absorption spectrum for the case of the strong coupling field in an $LG^{1}_{0}$ mode, as a function of the probe detuning $\Delta_{p}$ at $r=w_{LG}/27$ (a), $r=w_{LG}/9$ (b), $r=w_{LG}/3$ (c) and $r=w_{LG}$ (d). Other parameters are the same as those used in figure \ref{fig4.1}.}
\label{fig4.2}
\end{figure}

To understand the origin of the difference, we have recalculated the absorption spectrum for the case of the strong coupling field in an $LG^{1}_{0}$ mode at the different r. Analysis is indicated that the dominant contribution to the absorption spectrum arises from the smaller r. As is seen from figure \ref{fig4.2}, spectrum shows a relatively narrow peak for small r, while for the larger one, two smaller peaks appear in the spectrum resulting in the Autler-Townes doublet. It is clear that for the smaller intensities, a peak rises out of the spectrum and therefore Autler-Townes doublet structure cannot be exhibited. Thus we can expect that the laser profile will play a major role in establishing the Autler-Townes doublet structure.

\begin{figure}
\centering
\includegraphics[width=7 cm]{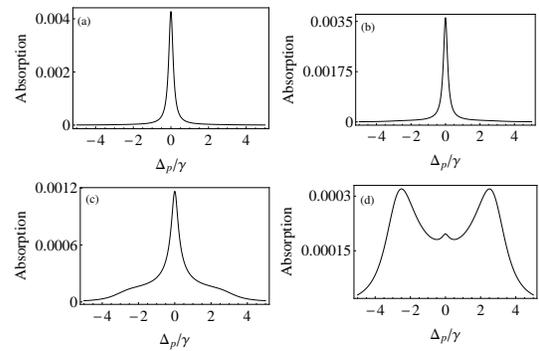}
\caption{Absorption spectrum as a function of the probe detuning $\Delta_{p}$ for the different $LG^{1}_{0}$ mode beam waist $w_{LG}= 135 \mu m$ (a), $w_{LG}= 270 \mu m$ (b), $w_{LG}= 540 \mu m$ (c) and $w_{LG}=1080 \mu m$ (d). Other parameters are the same as those used in figure \ref{fig4.1}.}
\label{fig4.3}
\end{figure}
With this background, we now discuss the effect of the LG mode beam waist on the linewidth of the absorption spectrum. Considering the different values of the beam waist, we find that for the small values of the beam waist, absorption spectrum does not exhibit the expected Autler-Townes doublet; instead, a prominent narrow central peak is seen, as shown in figure \ref{fig4.3}. Generally, it seems that the Autler-Townes doublet structure can hardly be created using the $LG^{1}_{0}$ mode.

\section{Double Dark Resonance}

Since the DDR does play a significant role in high resolution spectroscopy \cite{garva,chu}, we are interested in finding out how important  the profile laser is and how the spectrum can be changed by the different spatial distribution of the intensity profile. In this section, this novel spectral feature for the system considered in Figure \ref{figd} is investigated by applying the weak perturbing field and neglecting the decay on transition $ \vert 3\rangle-\vert 1\rangle$. Formation of a very sharp central peak caused by three-photon resonance in the spectrum is a characteristic feature of the phenomena \cite{1feng}.
\begin{figure}[h]
\centering
\includegraphics[width=7 cm]{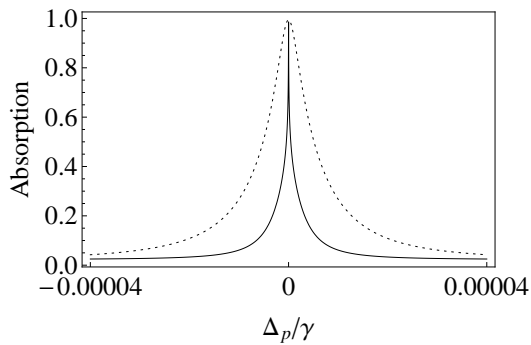}
\caption{ Normalized absorption spectrum as a function of the probe detuning $\Delta_{p}$ for the $LG^{1}_{0}$ (solid line) and Gaussian beams (dotted line) using $\gamma_{13}=0$ and $g_{41}=0.04 \, \gamma$. As the curves show double dark resonance is narrower for $LG^{1}_{0}$ mode than Gaussian mode. Other parameters are the same as those used in figure \ref{fig4.1}.}
\label{fig5}
\end{figure}

Figure \ref{fig5} shows the absorption spectrum versus the probe detuning $\Delta_{p}$ for both $LG^{1}_{0}$ and Gaussian fields, when we apply the weak perturbing field with Rabi frequency $g_{41}=0.04 \gamma$. The results show a narrow absorption peak at around zero detuning for both profiles. In particular, it shows that the three-photon absorption peak of double dark resonance is significantly narrower than that obtained with the Gaussian beam. For the above parameters, the calculated linewidth of the absorption peak for $LG^{1}_{0}$ and Gaussian profiles are about $1.8\times 10^{-6} \gamma$ and $2.1 \times 10^{-5} \gamma$, respectively. As the linewidth is approximately 12 times lower compared with Gaussian beam, it could improve super ultra high spectroscopy. This result will also open up possibilities for new applications such as metrology and laser isotope separation methods.

\section{Conclusion}

In conclusion, the influence of laser profile on the linewidth of the optical spectrum in the multi-photon resonance condition has been studied. First, we investigate the absorption spectrum for a two-level system and show that the Laguerre-Gaussian field results in a narrower peak than that of the Gaussian one. Then, the absorption in a two-level pump-probe atomic system is studied and it is shown that using the Laguerre-Gaussian fields, a narrower two-photon absorption peak is obtained compared to the usual Gaussian ones. We also study the effect of the laser profile on the coherent population trapping in the $\Lambda$-type molecular open systems. For the spatially-dependent Rabi frequencies, narrow peaks in the population of the levels appear in both profiles, yet the Laguerre-Gaussian fields can reduce the linewidth of the corresponding spectrum. In addition, we study the effect of the Doppler averaging on the electromagnetically induced transparency effect for a $\Lambda$-type atomic system. Comparing with the Gaussian field, the use of a Laguerre-Gaussian beam makes the linewidth of the spectrum narrower. Moreover, the influence of the laser profile on the Autler-Townes doublet structure in the absorption spectrum for a four-level atomic system is investigated. Considering the different values of the Laguerre-Gaussian mode beam waist, we find that for the small values of waist, the Autler-Townes doublet can be removed and the absorption spectrum reveals a prominent narrow central peak. Finally, the effect of the laser profile on the linewidth of the sub-natural three-photon absorption peak of double dark resonance in the four-level atomic system is studied and we demonstrate that the Laguerre-Gaussian beams induce the significant narrowing so that a linewidth about twelve times lower will be possible in this scheme. Since the Laguerre-Gaussian beams can reduce the linewidth of the all above-mentioned phenomena, the use of these beams are preferable for super ultra high resolution spectroscopy applications.

\end{document}